# Ultrafast Raman thermometry in driven YBa$_2$Cu$_3$O$_{6.48}$

T.-H. Chou[1], M. Först[1], M. Fechner[1], M. Henstridge[1], S. Roy[1], M. Buzzi[1], D. Nicoletti[1], Y. Liu[2], S. Nakata[2], B. Keimer[2], A. Cavalleri[1,3,*]

[1] Max Planck Institute for the Structure and Dynamics of Matter, 22761 Hamburg, Germany
[2] Max Planck Institute for Solid State Research, 70569 Stuttgart, Germany
[3] Department of Physics, Clarendon Laboratory, University of Oxford, Oxford OX1 3PU, United Kingdom

Signatures of photo-induced superconductivity have been reported in cuprate materials subjected to a coherent phonon drive. A "cold" superfluid was extracted from the transient Terahertz conductivity and was seen to coexist with "hot" uncondensed quasi-particles, a hallmark of a driven-dissipative system of which the interplay between coherent and incoherent responses are not well understood. Here, time resolved spontaneous Raman scattering was used to probe the lattice temperature in the photo-induced superconducting state of YBa$_2$Cu$_3$O$_{6.48}$. An increase in lattice temperature of approximately 80 K was observed by measuring the time dependent Raman scattering intensity of an undriven "spectator" phonon mode. This is to be compared with an estimated increase in quasi-particle temperatures of nearly 200 K. These temperature changes provide quantitative information on the nature of the driven state and its decay, and may provide a strategy to optimize this effect.

*andrea.cavalleri@mpsd.mpg.de

The equilibrium superconducting transition in cuprate superconductors is characterized by a number of features in the terahertz-frequency optical response [1-4]. For normal state YBa$_2$Cu$_3$O$_{6+x}$ (T > T$_C$), the real part of the optical conductivity $\sigma_1(\omega)$ exhibits a featureless spectrum with a finite value, which descends from the Drude response of quasi-particles at frequencies below the scattering rate ($\omega \ll 1/\tau$). Along the c-axis of YBa$_2$Cu$_3$O$_{6+x}$, where one finds a semiconducting normal state response, this feature reflects the population of thermally excited quasi-particles [5, 6]. Upon cooling into the superconducting state, the $\sigma_1(\omega)$ spectral weight in the frequency range below the superconducting energy gap is reduced, and a divergent 1/ω response in the imaginary part of the optical conductivity $\sigma_2(\omega)$ appears, indicating the onset of dissipation-less transport.

When mid-infrared pulses are used to resonantly drive apical oxygen phonons in the normal state (T > T$_c$) of underdoped YBa$_2$Cu$_3$O$_{6+x}$, changes in the terahertz optical properties are observed with the same 1/ω divergence in the transient $\sigma_2(\omega)$, connected to non-equilibrium superconducting-like properties [7-10]. Figure 1(a) and 1(b) display one example of these induced optical conductivity features, measured along the c-axis of YBa$_2$Cu$_3$O$_{6.48}$ at T = 100 K, showing the equilibrium state (gray line) and the light-induced state (red circles) [11].

The nature of this non-equilibrium state was recently studied in a series of complementary experiments, based on time-resolved measurements of the near-infrared reflectivity R(t) and the second-order nonlinear optical susceptibility χ$^{(2)}$(t). These studies revealed a number of coherently oscillating normal modes, including three coherent Raman-active phonons at frequencies of 3.7, 4.6, and 5.1 THz (Fig. 1d, gray [12-14]), two resonantly driven infrared-active phonons at 17.5 and 20 THz (Fig. 1d, orange), and, most importantly, two electronic modes at 2.5 and 14 THz (Fig. 1d, red). The two coherent electronic modes were assigned to amplified inter-bilayer and intra-bilayer Josephson plasmons [14]. The diagram in Fig. 1e summarizes the current understanding of these coherent couplings [14, 15].

The present study focuses on *incoherent* energy transfer in driven YBa$_2$Cu$_3$O$_{6.48}$. Firstly, the mid-IR excitation pulses not only drive coherent phonons but also excite charge carriers, which in turn couple to other modes of the system, including phonons, spin excitations, and plasma modes (see Fig. 2a). Figure 2a depicts schematically the expected quasi-particle heating, represented by the corresponding electron density of states (eDOS) and population n(E). Secondly, one expects that the coherently driven phonons will also decay into incoherent vibrational modes through spontaneous ionic Raman processes, which will raise the temperature of many low-energy degrees of freedom. As an example of these dissipative dynamics, we focus here on a high-frequency Raman mode at 15 THz [16, 17], whose period is too high to be excited coherently by the mid-infrared pump pulses [12, 18].

To estimate the incoherent response of hot quasi-particles, we tracked the $\sigma_1(\omega)$ conductivity spectra at different time delays after the optical excitation. Following the same analysis as reported in Ref. [11], we calculated the $\sigma_1(\omega)$ spectral weight between 0.6 and 2.25 THz (20 – 75 cm$^{-1}$) as $\int_{0.6\,THz}^{2.25\,THz} \sigma_1(\omega) d\omega$ to determine the normal fluid density. Then, we compared the value of this integral to its equilibrium values taken at different temperatures to infer an "equivalent temperature" of the quasi-particles. The time evolution of these transient spectral weights (Fig. 2c) shows that the quasi-particles undergo a temperature increase $\Delta T$ of 200 K and decay back to the equilibrium temperature in about 5 ps. Although this estimate is by no means a precise measurement of the quasi-particle temperature, it provides a useful point of reference.

Here, we complement the measurements reported above with time-resolved spontaneous Raman scattering experiments [19-27] to probe the lattice temperature under the same mid-infrared excitation condition. For these experiments, we focused on measurements of a "spectator mode", not excited directly by the pump and with frequencies that were high enough to prevent their coherent excitation (only possible for phonon modes with full periods longer than twice the pump pulse duration).

Figure 3a shows a schematic of our time-resolved Raman scattering setup. The $YBa_2Cu_3O_{6.48}$ crystal was excited by 800-fs mid-infrared pulses, centered at 20 THz frequency and polarized along the crystal c-axis to resonantly drive the infrared-active apical oxygen phonons that induce the transient superconducting state. Probe pulses with 405 nm wavelength, 600 fs time duration, and 100 cm$^{-1}$ bandwidth were used for the Raman scattering process. The probe photons, scattered from the sample, were analyzed by a spectrometer equipped with a thermoelectric-cooled CCD camera. All Raman spectra presented in this paper were acquired in the backscattering geometry, with both the input probe pulses and the scattered photons polarized along the $YBa_2Cu_3O_{6.48}$ c-axis.

Figure 3b displays the Raman spectra obtained at equilibrium and at a time delay t = 0.1 ps after excitation. We find the Stokes (S) and anti-Stokes (AS) peaks of the apical oxygen $A_{1g}$ phonon at ± 15 THz, which are offset in frequency from the coherently driven infrared-active $B_{1u}$ symmetry mode around 20 THz. The intensity ratio of the AS and S peaks, $I_{AS}/I_S$, is related to the distribution of phonon populations in the ground state and the first excited state and allows to calculate the temperature of this Raman mode using [28]

$$\frac{I_{AS}}{I_S} = \frac{(\omega_0+\omega_{ph})^3}{(\omega_0-\omega_{ph})^3} \frac{n}{n+1} = \frac{(\omega_0+\omega_{ph})^3}{(\omega_0-\omega_{ph})^3} e^{-\hbar\omega_{ph}/k_B T_{ph}}. \quad (1)$$

Here, $\omega_0$ is the central frequency of probe pulses, $\omega_{ph}$ is the frequency of the Raman mode, $k_B$ is the Boltzmann constant, and $T_{ph}$ is the phonon temperature defined by the occupation number $n = (e^{\hbar\omega_{ph}/k_B T_{ph}} - 1)^{-1}$ based on the Bose-Einstein distribution. The cubic term in Eq. 1 arises from the frequency dependence of the Raman scattering cross section $\sigma_R \propto \omega^4$. Note that the photon counting detector used here, for which the read-out signal strength is proportional to the total photon number rather than the total energy, requires to use a cubic instead of a quartic term [28].

The time-averaged probe power was kept below 0.2 mW to minimize heating caused by the probe pulse itself [22, 26]. As supporting evidence, the equilibrium Raman spectrum (Fig. 3b, gray line) measured at room temperature demonstrates an intensity ratio $I_{AS}/I_S$ of ~0.11, corresponding to a phonon temperature $T_{ph}$ of ~300 K. With mid-IR excitation, the pumped Raman spectrum taken at time delay t = 0.1 ps (Fig. 3b, orange line) shows an enhancement in the amplitude of the AS peak, while the S peak remains unchanged, indicating an increase of phonon temperature after photo-excitation.

Figure 3c shows the time evolutions of the $A_{1g}$ phonon temperature in $YBa_2Cu_3O_{6.48}$ measured at cryostat temperatures $T_{cryo}$ of 295 and 100 K. Both transients display a prompt rise in phonon temperature near time zero followed by an exponential decay to a finite temperature, which can be interpreted as the thermalization of this phonon with other degrees of freedom in the system. Note that for the low-temperature experiments, the measured $T_{ph}$ = 179 K at negative time delays does not correspond the cryostat temperature $T_{cryo}$ = 100 K. We attribute this issue to average heating induced by the probe pulse itself [22, 26], which is difficult to avoid due to the low heat capacity in $YBa_2Cu_3O_{6.48}$ at 100 K. Note also that at these low temperatures the AS peak for a 15 THz phonon is barely detectable. The amplitude of the AS peak at $T_{ph}$ = 100 K is 100 times smaller than in the $T_{ph}$ = 300 K data (Fig. 3b, gray line). This adds some uncertainty in the determination of the base temperature of YBa$_2$Cu$_3$O$_{6.48}$ near 100 K.

We first focus on the data at long-time delays (>5 ps), shown in Figure 3c. Here, we expect the different degrees of freedom in the material to have reached thermal equilibrium, and we compare the measured temperature change to a calculation based on the total energy absorbed by the sample, the excited volume, and the heat capacity (see Supplemental Material S4 for details). For a mid-infrared excitation density of 8 mJ/cm², shown as the red dashed line in Fig. 3c, the calculated temperature of the YBa$_2$Cu$_3$O$_{6.48}$ surface increased from $T_i$ = 295 K (179 K) and thermalized at $T_f$ = 310 K (195.4 K). The measured phonon temperatures at negative and long positive time delay agree with the calculation.

We next turn to the results at earlier time delays, that is before thermalization. A non-equilibrium phonon temperature of 370 K (260 K) was measured at the peak of the signal for the initial temperature of 295 K (179 K), well above the equilibrated temperature of 310 K (195.4 K) at longer delays. This peak temperature is far lower than what reported for near-infrared excitation in Reference [26]. Also, this lattice temperature increase is smaller than the estimated quasi-particle temperature extracted from the measurements of the optical conductivity $\sigma_1(\omega, t)$, from which we estimated a transient carrier temperature increase of 200 K, and a similar decay time constant of ~1.5 ps.

Given that our time-resolved Raman experiment might underestimate the peak lattice temperature due to the 600-fs probe pulse, we conducted a deconvolution analysis on the Raman data (Fig. 3c) to achieve a more precise estimation (see Supplemental Materials S3).

The results indicate that the maximum increase in lattice temperature is approximately 80–85 K, still lower than that of the estimated quasi-particle temperature of 200 K.

These incoherent "hot" degrees of freedom are to be compared to the dynamics of the superfluid, manifested in the strength of the divergent imaginary part of the optical conductivity $\sigma_2(\omega)$. We conducted the analysis on the $\sigma_2(\omega)$ spectra (reported in Ref. [11]) at different time delays and calculated the values of $\lim_{\omega \to 0} \omega \sigma_2(\omega)$, which is directly related to superfluid density. By comparing these values with equilibrium optical conductivity spectra taken at multiple temperatures below $T_C$, we were able to extract the "equivalent temperature" of the superfluid. The time evolution of the coherent superfluid in response to the mid-infrared excitation is presented in Figure 4b. In contrast to the longer-lived heating effects observed on incoherent phonons and quasi-particles, the equivalent "cooling" of superfluid only persists within the duration of pump pulse.

As the quasi-particles are at all times hotter than the lattice, it is likely that these determine the dissipation dynamics. At least two scenarios emerge at this stage. A stronger superfluid vis-a-vis hot carriers and only a moderately heated lattice may connect the present physics to the well-known microwave enhanced superconductivity mechanism [29]. However, this appears unlikely in the present context as the effect is far too large to be explained by the Eliashberg mechanism [30] and the observed response is resonant with the optical phonon drive. A second scenario, which we believe is more likely, would be one in which the hot quasi-particles are detrimental to the coherent state, whichever mechanism one may invoke. In this picture, the decoherence time of the light induced superconducting-like state is probably determined by the decay of hot quasi-particles and not by the hot lattice. As the lifetime of the coherent phase is never observed to be longer than the optical pulse [11], one should focus on optimizing the conditions of the drive to minimize the excitation of hot carriers. Suitably shaped pump pulses or trains of pump pulses may provide a stronger and longer-lived photo-induced state.

We acknowledge support from the Deutsche Forschungsgemeinschaft (DFG; German Research Foundation) via the excellence cluster "CUI: Advanced Imaging of Matter" (EXC 2056, project ID 390715994) and the priority program SFB925 (Project ID 170620586).

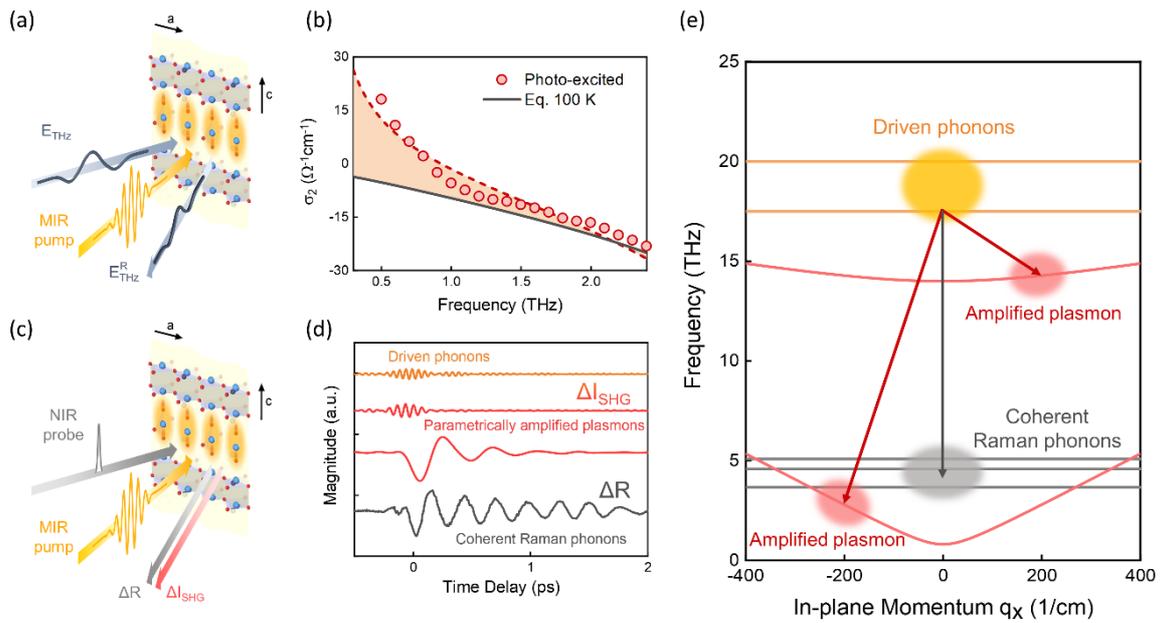

**Figure 1 Coherent couplings in optically driven YBa$_2$Cu$_3$O$_{6.48}$.** **(a)** Schematic of time-resolved terahertz spectroscopy. The mid-IR pump pulse (yellow) resonantly drives IR-active c-axis apical oxygen phonons in YBa$_2$Cu$_3$O$_{6.48}$. Single-cycle terahertz probe pulses (slate gray) were reflected from the sample and detected by phase-sensitive electro-optic sampling. **(b)** Imaginary part of the c-axis optical conductivity $\sigma_2(\omega)$, measured in YBa$_2$Cu$_3$O$_{6.48}$ at T = 100 K. Gray line and red circles show the equilibrium and transient $\sigma_2(\omega)$, respectively. The red dashed line is a fit to the transient spectrum[11] **(c)** Schematic of time-resolved optical spectroscopy. The pump pulse (yellow) is the same as in panel (a). Near-IR probe pulses (gray) were detected in reflection geometry. Gray and red arrows represent reflectivity changes at the fundamental frequency ($\Delta R$) and intensity changes of second harmonic generation ($\Delta I_{SHG}$), respectively. **(d)** $\Delta R$ (gray) shows an oscillatory response at the Raman phonon frequencies. This datum was employed from Ref. [14] Fig. 1b with the slowly varying background subtracted. $\Delta I_{SHG}$ shows instead coherent oscillations of the driven infrared-active phonons (orange) and parametrically amplified intra-bilayer and inter-bilayer Josephson plasma polaritons (red). These data were utilized from Ref. [14] Fig. 2e and applied FFT band pass filter with frequency windows of 0–5, 12–17, and 17–22 THz. **(e)** Frequency-momentum diagram showing the dispersion curves of driven infrared phonons (orange), Josephson plasmons (light red), and Raman phonons (light gray). The dark gray arrow shows the coupling pathway of stimulated ionic Raman processes between the driven phonons and Raman phonons. The dark red arrows indicate instead the three-mode mixing process between driven phonons and Josephson plasmons.

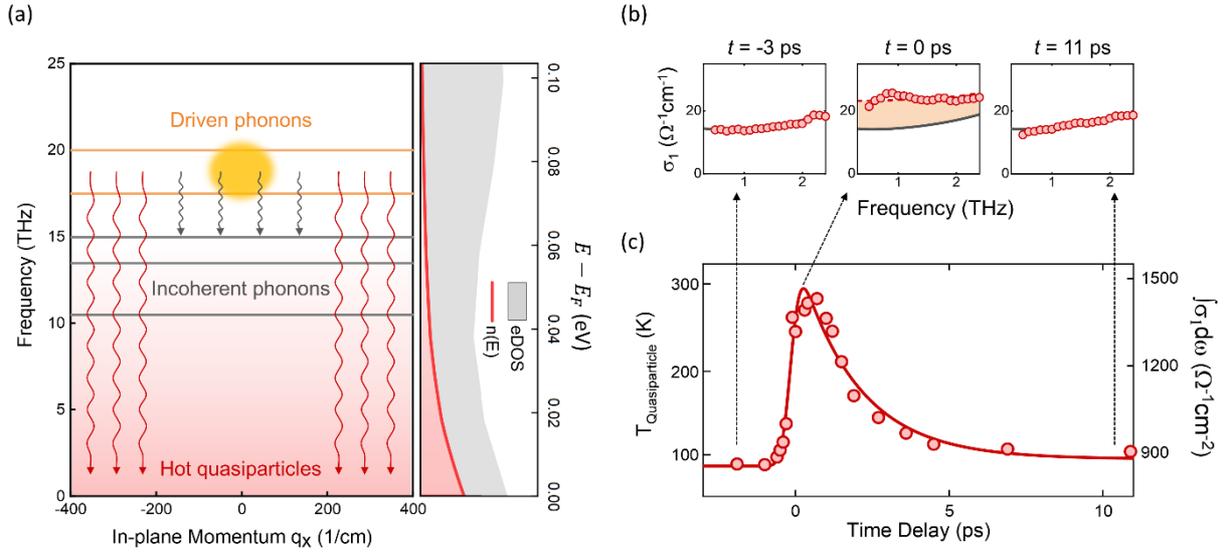

**Figure 2: Incoherent couplings in optically driven YBa$_2$Cu$_3$O$_{6.48}$.** (a) (left panel) Frequency-momentum diagram with the dispersion curves of resonantly driven infrared phonons (orange), and incoherent Raman phonons at 10.5, 13.5, and 15 THz[16, 17] (light gray). The red gradient background represents the electron distribution above the Fermi energy, which is at zero frequency in this diagram. The wiggling arrows indicate the incoherent energy transfer to Raman phonons (dark gray) and quasi-particles (dark red). Right panel: Electron density of state *eDOS* (gray) and population *n(E)* (red). The former was obtained by ab-initio calculations, while the latter was calculated by multiplying the *eDOS* with a Fermi-Dirac distribution at *T* = 100 K. (b) Real part of the c-axis optical conductivity, σ$_1$(ω), of YBa$_2$Cu$_3$O$_{6.48}$ measured at time delays of *t* = -3, 0, and +11 ps. In each panel, gray line and red circles represent the equilibrium and transient σ$_1$(ω) at T = 100 K, respectively. The red dashed line is a fit to the transient spectrum[11]. (c) Equivalent quasi-particle temperature as a function of time delay. The red line is a fit with an error function and an exponential decay. The quasi-particle temperature was determined from the integrated spectral weight $\int_{0.6\,THz}^{2.25\,THz} \sigma_1(\omega)d\omega$. The corresponding spectral weight values are reported on the right axis.

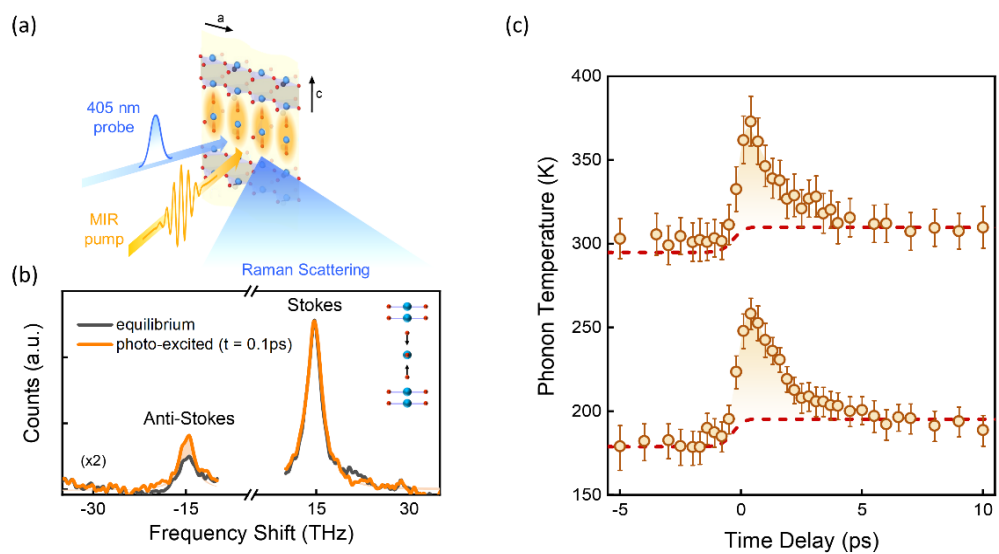

**Figure 3: (a)** Schematic of time-resolved spontaneous Raman spectroscopy. The pump (yellow) is the same as in Fig. 1. 405-nm probe pulses (blue) were used for the Raman scattering process. The scattered photons were spectrally analyzed. **(b)** The Raman spectra were measured at T = 295 K at equilibrium (black) and after photoexcitation (orange). The displayed data were obtained by subtracting background spectra and high-order side peaks from the raw data (see Supplemental Materials S2). The peaks at ± 15 THz represent the Stokes (+) and Anti-Stokes (−) responses of the $A_g$ symmetry apical oxygen Raman mode. Inset: atomic motions along the c-axis apical oxygen Raman-active mode coordinates. **(c)** Phonon temperature as a function of time delay. These data were measured at cryostat base temperatures of 295 K (upper) and 100 K (lower). The dashed red lines were obtained from heating calculation (see Supplemental Materials S4).

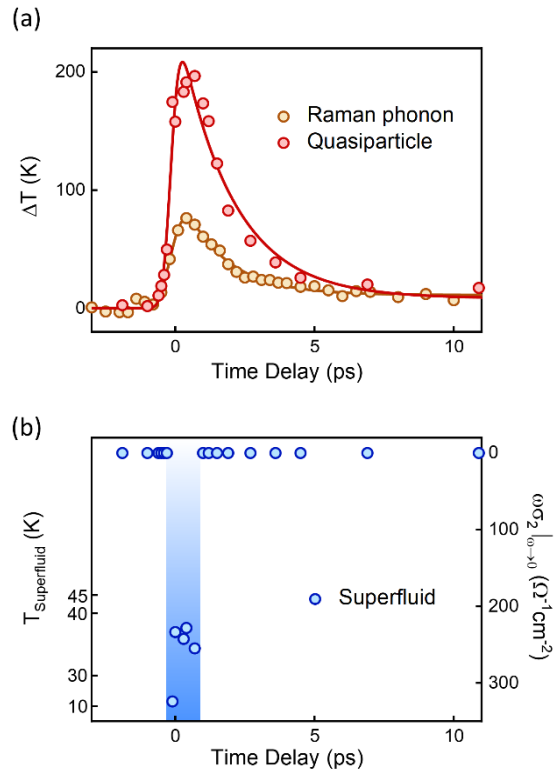

**Figure 4: (a)** Temperature changes of quasi-particles and Raman phonons as a function of time delay. The solid lines are fits to the data with a model including an error function and an exponential decay. **(b)** Equivalent temperatures of the superfluid as a function of time delay[11]. The blue shaded area highlights the time delay window within which light-induced superconductivity was observed. The equivalent temperature of the superfluid was determined from the coherent superconducting-like response $\lim_{\omega \to 0} \omega\sigma_2(\omega)$. The corresponding values are reported on the right axis.